\documentclass[preprint,aps,showpacs,superscriptaddress,nofootinbib]{revtex4-1}
\usepackage{graphicx}
\usepackage{multirow}
\usepackage[caption=false]{subfig}
\usepackage{dcolumn}
\usepackage{amsfonts, amssymb,amsmath}
\newcommand{\Lam}{\Lambda(1405)}
\begin{document}

\title{Lessons from fitting the lowest order energy independent
  chiral based $\bar{K}N$ potential to experimental data.}

\author{J. R\'{e}vai}
\affiliation{Wigner Research Center for Physics, RMI,
H-1525 Budapest, P.O.B. 49, Hungary}

\email{revai.janos@wigner.hu}

\date{\today}

\begin{abstract}

It is shown, that fitting parameters of a $\bar{K}N$ interaction model to different sets of experimental data can lead to physical conclusions which might provide a deeper insight into the physics of this multichannel system. The available experimental data are divided into three parts: the "classical" set consisting of the low-energy $K^-p$ cross sections and the threshold branching ratios, the SIDDHARTA $1s$ level shift in kaonic hydrogen and the CLAS photoproduction data. We have fitted the parameters of the potential to different combinations of these data. We found, that  the two poles corresponding to the $I=0$  nuclear quasi-bound state ($\Lam$) and to the $K^-p$ $1s$ atomic level seem to resist to their simultaneous reproduction at the right place, though a more or less satisfactory compromise can be achieved. Potentials with the $\Lam$ pole pinned down close to the PDG value fail to reproduce the classical two-body data with an acceptable accuracy. We also added comments on two papers criticizing the potential used in the fits.

\end{abstract}

%13.75.Jz: kaon-baryon interactions
%36.10.Gv: Mesonic atoms and molecules, hyperonic atoms and molecules
\maketitle

%%%%%%%%%%%%%%%%%%%%%%%%%%%%%%%%%%%%%%%%%%%%%%%%%%%%%%%%%%%%%%%%
\section{Introduction}
\label{Introduction_sect}
In the last decade the antikaon nucleon interactions attracted considerable
attention. Several models were proposed and at present it is generally believed,
that the correct theoretical background for derivation of this quantity
is the chiral perturbation expansion of the SU(3) meson-baryon interaction
Lagrangian. The abundant literature on this subject is summarized in several
recent review papers, e.g. \cite{Mai,Kamiya,Cieply} and references therein.
We do not seek here after repeating this uneasy task.
There are two main directions along which these interactions are constructed:
\begin{itemize}
\item[a)] A possibly complete reproduction of multichannel two-body data in a wide
  energy range. These approaches use relativistic formulation and in order
  to obtain good agreement with experimental data, they go beyond the lowest
  order term in the chiral perturbation expansion. While the resulting interactions
  provide a good description of meson-baryon two-body data, they are
  not suited to be used for the theoretical description of systems involving
  more than two particles.

\item[b)]  In view of possible existence of kaonic nuclear clusters, the main idea of the
  other approach is to construct a potential, which can be used for calculation
  of $n>2$ systems. In this case the simplicity of the potential is essential, therefore
  the usual practice is to keep the form of the lowest order Weinberg-Tomozawa (WT) term of the chiral
  expansion and try to adjust it to the experimental data. For this purpose data are
  selected from the vicinity of $\bar{K}N$ threshold and in general from channels,
  open at this energy. Since the $n>2$ calculations at present can be performed only
  in a non-relativistic, potential based framework, for consistence this formalism
  has to be kept also at the potential fitting level.
\end{itemize}

Whatever potential model is chosen, at a certain point its parameters have to be fitted to
a set of experimental data. This process is usually not reported in detail when a
new potential is introduced. We think, however, that the fitting procedure might reveal
some important issues providing a deeper insight into the physics of the $\bar{K}N$
system. In the present paper we try to demonstrate this idea on the example of a recently
introduced $\bar{K}N$ interaction, belonging to the group b) \cite{revai1,revai2}. The potential is energy-independent
and leads to a one-pole structure of the $\Lambda(1405)$ resonance, in contrast to the two-pole
one, obtained from the commonly used chiral based $\bar{K}N$ potentials.%\footnote{Here it seems to be
%appropriate to mention, that a recent thorough analysis of all relevant experimental data
%\cite{hyperon} found no evidence for the existence of the second pole.}

\section{Input: theoretical model and experimental data}

The potential which we shall use to demonstrate the effects of data fitting  \cite{revai1,revai2},  is
an energy-independent implementation of the lowest order  WT term of the chiral $SU(3)$ meson-baryon interaction
Lagrangian, designed for non-relativistic calculations in two- and few-body systems. Its actual form is a two-term multichannel separable
potential:
\begin{equation}
\langle k_i|V_{ij}|k_j\rangle = \lambda_{ij}(g_{iA}(k_i)g_{jB}(k_j)+g_{iB}(k_i)g_{jA}(k_j)),
\end{equation}
where  the channel indices $i(j)=1,2,3,4,5$ correspond to the channels $[\bar{K}N]^{I=0},[\bar{K}N]^{I=1}$,\ $[\Sigma\pi]^{I=0},[\Sigma\pi]^{I=1},[\Lambda\pi]^{I=1}$ , respectively. The form factors $g_{iA}$ and
$g_{iB}$ are defined as
\begin{equation}
g_{iA}(k_i)=\left({\beta_i^2 \over \beta_i^2+k_i^2}\right)^2; \qquad g_{iB}(k_i)=g_{iA}(k_i)
\left(m_i+{k_i^2 \over 2\mu_i}\right),
\end{equation}
where $k_i$ is the relative momentum, $m_i$ and $\mu_i$ are the meson- and the reduced masses
in channel $i$, while the $\beta_i$ are the corresponding range (or cut-off) parameters. The coupling constant $\lambda_{ij}$
has the form
\begin{equation}
\lambda_{ij}=-{c_{ij} \over 64 \pi^3 F_iF_j\sqrt{m_im_j}},
\end{equation}
with $c_{ij}$ being the $SU(3)$ Clebsh-Gordan coefficients, while $F_i(F_j)$ stand for the meson decay constants in channel $i(j)$.
The way, how to calculate two-body observables from this interaction is described in detail in \cite{revai1,revai2}. The
potential contains 7 adjustable parameters: the two meson decay constants $F_{\bar{K}}$ and $F_{\pi}$ and the five range parameters
$\beta_i$ in channels $i=1,2,3,4,5$ .

There are three types of data, to which $\bar{K}N$ potentials are usually fitted. The "classical" group
consists of the six old elastic and inelastic low-energy $K^-p$ cross sections with a rather poor accuracy
and the three threshold branching ratios $\gamma,R_n\ {\rm and}\ R_c$:
$$
 \gamma={\sigma(K^-p\to\Sigma^-\pi^+)\over \sigma(K^-p\to\Sigma^+\pi^-)}\ R_n={\sigma(K^-p\to \pi^0\Lambda)\over
 \sigma(K^-p\to \pi^0\Lambda, \pi^0\Sigma^0 )}\
 R_c={ \sigma(K^-p\to\Sigma^-\pi^+,\Sigma^+\pi^- )\over \sigma(K^-p\to{\rm all\ inelastic\ channels})}
$$
with usual experimental values
$$
\gamma=2.36\pm0.04;\quad R_n=0.189\pm0.015;\quad R_c=0.664\pm0.011\ .
$$
In order to avoid overweighting $\gamma$ in $\chi^2$ we adopt the reasoning of Guo and Oller \cite{Guo}
and take an experimental value with a somewhat increased error: $\gamma=2.36\pm0.1$.
The experimental values for the ``classical data'' are taken from the set of papers \cite{expref}. 

A "modern" piece of data is the recently measured complex energy shift $\Delta E$ of the $1s$ atomic level in kaonic hydrogen \cite{SIDD}:
$$
 \Delta E=283\pm36({\rm stat})\pm 6({\rm syst})\ -(271\pm 45({\rm stat})\pm 11({\rm syst}))i\ eV
$$
There is some confusion in the literature how to treat the statistical and systematic errors in the fitting procedure. We have added them quadratically and used
$$
\Delta E=283\pm36-(271\pm46)i\ eV
$$
And, finally, a somewhat controversial experimental information is provided by the PDG value for the position of the $\Lambda(1405)$ resonance ($\Lambda^*$ in the following):
$$
E(\Lambda^*)=1405^{+1.3}_{-1.0}-(25\pm1)i\ MeV
$$
We call it controversial, since it can be seen only in different reactions involving more than two particles. As a consequence, the position and shape can (and do) depend on the reaction mechanism and details. Therefore, in our opinion, the direct identification of the PDG resonance parameters, deduced from observed line shapes, with a pole position of the two-body $\bar{K}N$ interaction is not justified. This practice leads to potentials with a strong $\bar{K}N$ attraction and deeply bound $\bar{K}N$ pairs, which in turn stimulated far reaching speculations about a new type, extra dense  kaonic nuclear matter.
\section{Fitting}
For our $\chi^2$ per degree of freedom we take the one used by the majority of papers dealing with this problem, 
the first one taking into account both cross sections and discrete data being probably the Borasoy, Nissler, Meissner paper  \cite{Bora},
 which tries to establish a certain balance between the
weights of discrete observables and cross section data:
\begin{equation}
\chi^2_{\rm d.o.f}= {\sum_k n_k \over K (\sum_k n_k-n_{P})}\sum_{k=1}^K{\chi^2_k \over n_k};\quad
\chi^2_k=\sum_{i=1}^{n_k}{(y_{k;i}^{\rm th}-y_{k;i}^{\rm exp})^2 \over \sigma_{k;i}^2},
\end{equation}
where $K$ is the number of different measurements included into the fit, $n_k$ stands for the number of data points in the $k$th measurement, $n_P$ corresponds to the number of free parameters (7 in our case) and $y_{k;i}^{\rm exp}(y_{k;i}^{\rm th})$ represent the $i$th experimental(theoretical) point in the $k$th data set with standard error $\sigma_{k;i}.$\footnote{The subscript d.o.f of $\chi^2$
will be omitted in the following}

We have performed several fits to different combinations of the three types of experimental data.
In general, the fits do not produce sharp minima: small changes in the parameters in the vicinity of the best values lead to other $\chi^2$ local minima, slightly larger, than the "best" ones. To make clearer the  contribution of different data to the resulting $\chi^2$ we divided it into two parts:
$$
\chi^2=\chi^2_{disc}+\chi^2_{cs}\ ,
$$
corresponding to the discrete and cross section data. Generally, the $\chi^2_{cs}> 0.7-0.8$ for the best fits and provide the dominant part of the total $\chi^2$. This is due to the large errors and spreading of the cross section data and cannot be substantially reduced by the fitting procedure.

\begin{figure}
  \centering
  \subfloat[Fit to classical cross sections .]{
    \includegraphics[width=0.485\linewidth]{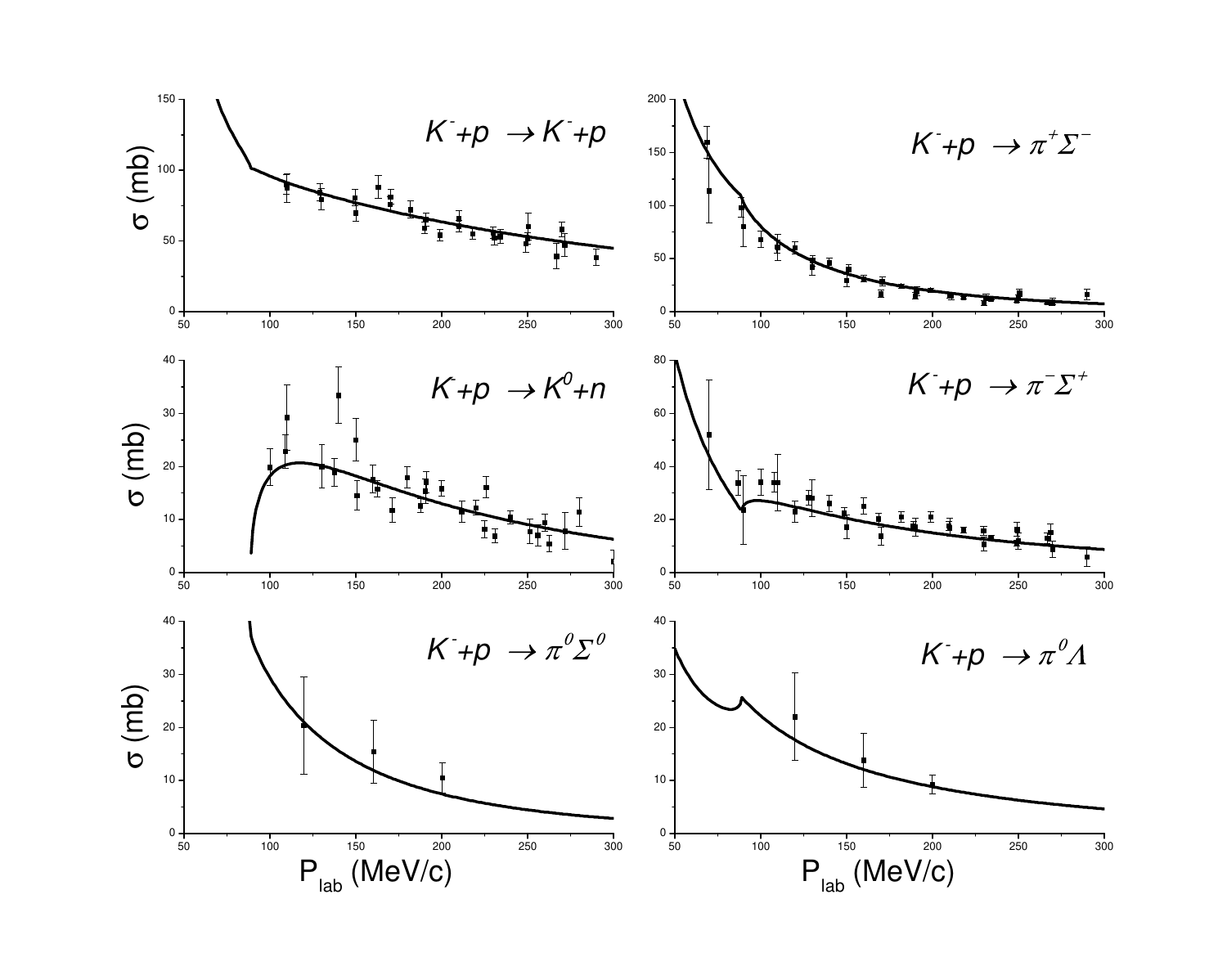}}
  \subfloat[Fit to CLAS data]{
    \includegraphics[width=0.485\linewidth]{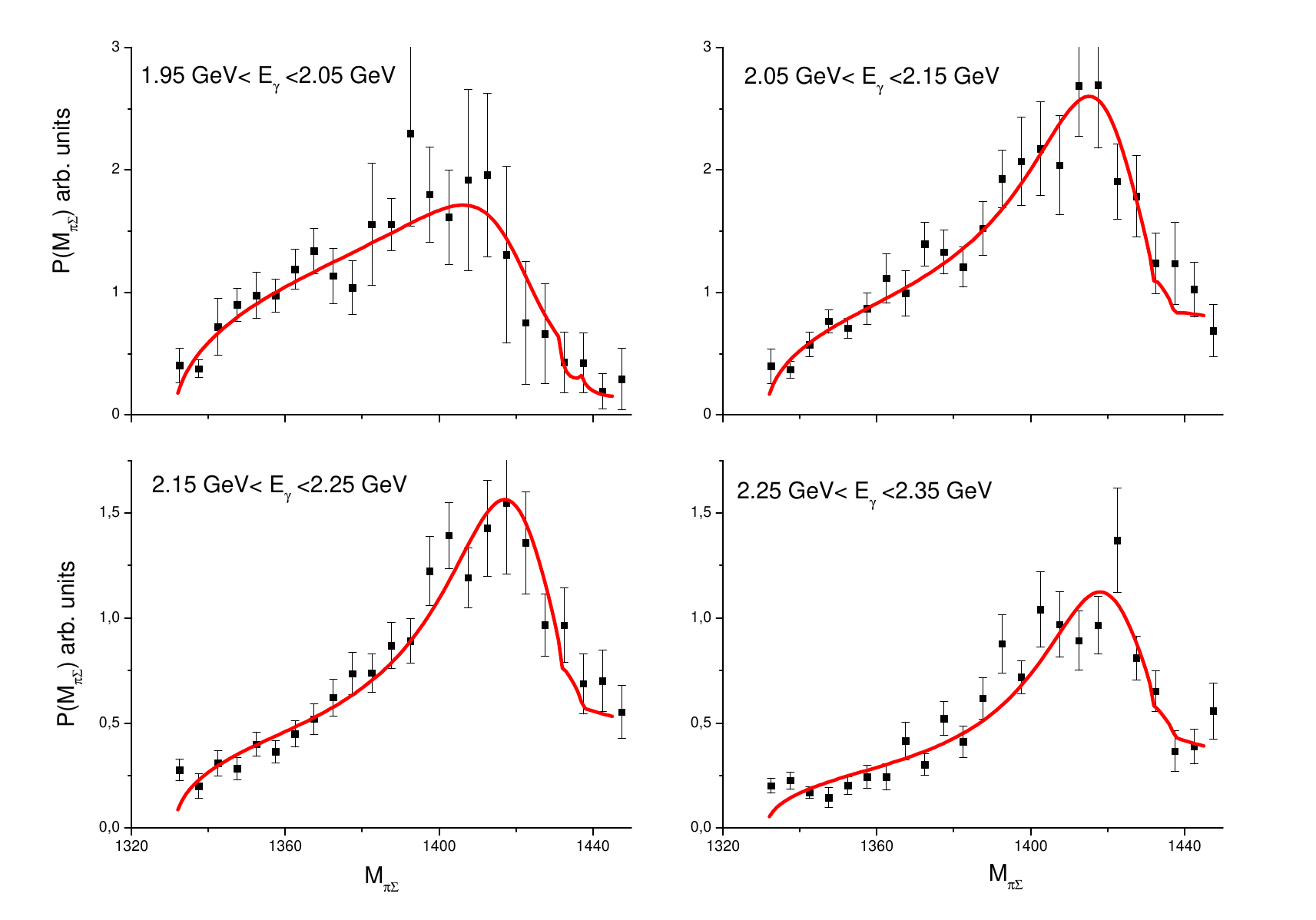}}
  \caption{Results of fit \bf{A}}
  \label{A}
\end{figure}

In our first fit ({\bf A}) we fitted the potential parameters to the "classical" data set. The results can be seen in the first column of the summary Tables \ref{sum} and \ref{param} and in part a) of Fig.\ref{A}. As for the quality of the fit, the discrete data (branching ratios) are within the experimental errors and the cross sections are also reproduced in a (visually) acceptable way. It can be seen, as we mentioned before, that the dominant part of the $\chi^2$ comes from $\chi^2_{cs}$.

\begin{table}
\begin{center}
    \begin{tabular}{c|c|c|c|c|c|c}
    \multicolumn{1}{c}{}&\multicolumn{5}{c}{\textbf{Fits}}&\\[3pt]
     \multicolumn{1}{c|}{}& \textbf{A} & \textbf{B} & \textbf{C}& \textbf{D} & \textbf{E}&\\
     \cline {1-6}
     $\chi ^2_{disc}$&0.074&0.209&0.357&0.58&1.50&\\
    $\chi ^2_{cs}$&0.841&0.931&0.700&1.46&1.28&\\
    $\chi ^2$&0.915&1.140&1.057&2.04&2.78&\\[2pt]
  \cline{2-7}
%\multicolumn{7}{c|}{}\\[0.5pt]
   \multicolumn{1}{c|}{}& \multicolumn{1}{c}{}&\multicolumn{3}{c}{Calculated values}& \multicolumn{1}{c|}{}&\multicolumn{1}{c|}{Exp}\\
    \hline
    $\gamma$ & 2.34 & 2.35 &2.34&2.34&2.35&\multicolumn{1}{c|}{2.36 $\pm$ 0.1}\\
    $R_c$ &0.671 & 0.669 & 0.675&0.685&0.685&\multicolumn{1}{c|}{0.664 $\pm$ 0.011}\\
    $R_n$  &0.195 & 0.199 &0.201&0.203&0.204&\multicolumn{1}{c|}{0.189 $\pm$ 0.015}\\
   Re($\Delta E)\ (eV)$&392 &277&329&369&352&\multicolumn{1}{c|}{283 $\pm$ 36}\\
 - Im($\Delta E)\ (eV)$& 232&329&275&137&143&\multicolumn{1}{c|}{271 $\pm$ 46}\\
Re($E(\Lambda^*))\ (MeV)$&1422&1440&1428&1407&1408&\multicolumn{1}{c|}{}\\
-Im($E(\Lambda^*))\ (MeV)$&20&27&24&23&25&\multicolumn{1}{c|}{}\\
    \hline
  \end{tabular}
  \end{center}
  \caption{Summary table of the fit results \textbf{A}-\textbf{E}}
      \label{sum}
\end{table}

\begin{table}
\begin{center}
    \begin{tabular}{ c|c|c|c|c|c|}
    \multicolumn{1}{c}{}&\multicolumn{5}{c}{\textbf{Fits}}\\[3pt]
     \multicolumn{1}{c|}{}& \textbf{A}& \textbf{B} & \textbf{C}&\textbf{D} & \textbf{E}\\
     \cline {1-6}
     $f_{\pi}$&62.8&117&63.5&80.6&71.4\\
    $f_{K}$&113&113&120&100&107\\
    $\beta_1$&945&853&968&888&948\\
    $\beta_2$&1346&981&1331&1233&1357\\
    $\beta_3$&412&547&372&513&465\\
    $\beta_4$&310&554&280&470&395\\
    $\beta_5$&236&406&214&323&267\\
    \hline
  \end{tabular}
  \end{center}
  \caption{Potential parameters obtained in the fits {\bf A} - {\bf E}. All values are in $MeV$.}
  \label{param}
\end{table}

The most remarkable feature of fit {\bf A} to the "classical" data is, that the position of the pole corresponding to $\Lambda^*$ is
\begin{equation}
E_{\bf A}(\Lambda^*)=1422-20i\ MeV,
\label{Lam*}
\end{equation}
while for the $1s$ level shift in kaonic hydrogen $\Delta E$ we have
$$
\Delta E_{\bf A}=392 - 232i\ eV,
$$
the real part of which is well out of the experimentally allowed range. These numbers suggest, that the simplest (WT) term of a chiral $SU(3)$ based $\bar{K}N$ interaction, when fitted to the "classical" data set, supports a $\Lambda^*$ pole at the position (\ref{Lam*}),while the same potential strongly overestimates the real part of $\Delta E$. The latter observation was already noticed in earlier fits to experimental data using chiral based potentials. Acceptable reproduction of the experimental $\Delta E$ could be achieved only by adding next order terms to the lowest order WT one \cite{Ikeda}.

The next question is, what happens to the fits if we try to force our simple potential to reproduce $\Delta E$, too? This is our fit {\bf B}, when we added to the "classical" data the experimental value of $\Delta E$.
The results are shown in column {\bf B} of Tables I and II and in part a) of Fig. 2. It can be seen, that the about 25\% worsening of $\chi^2$ comes essentially from the discrete part, signifying, that the WT based potential "has difficulties" in reproducing $\Delta E$. The real part of the level shift $\Delta E_{\bf B}$ is
close to the experimental value, while its imaginary part is still somewhat outside of the experimentally allowed range. However, the price for bringing the real part close to the experiment is that the $\Lambda^*$ pole moved above the $\bar{K}N$ threshold:
$$
E(\Lambda^*)_{\bf B}= 1440-27i\ MeV,
$$
what is probably unacceptable from experimental point of view. This is confirmed by by our fits to the CLAS $\Lambda(1405)$ data (Fig.2, part b)), which we shall discuss later.

\begin{figure}
  \centering
  \subfloat[Fit to classical cross sections .]{
    \includegraphics[width=0.485\linewidth]{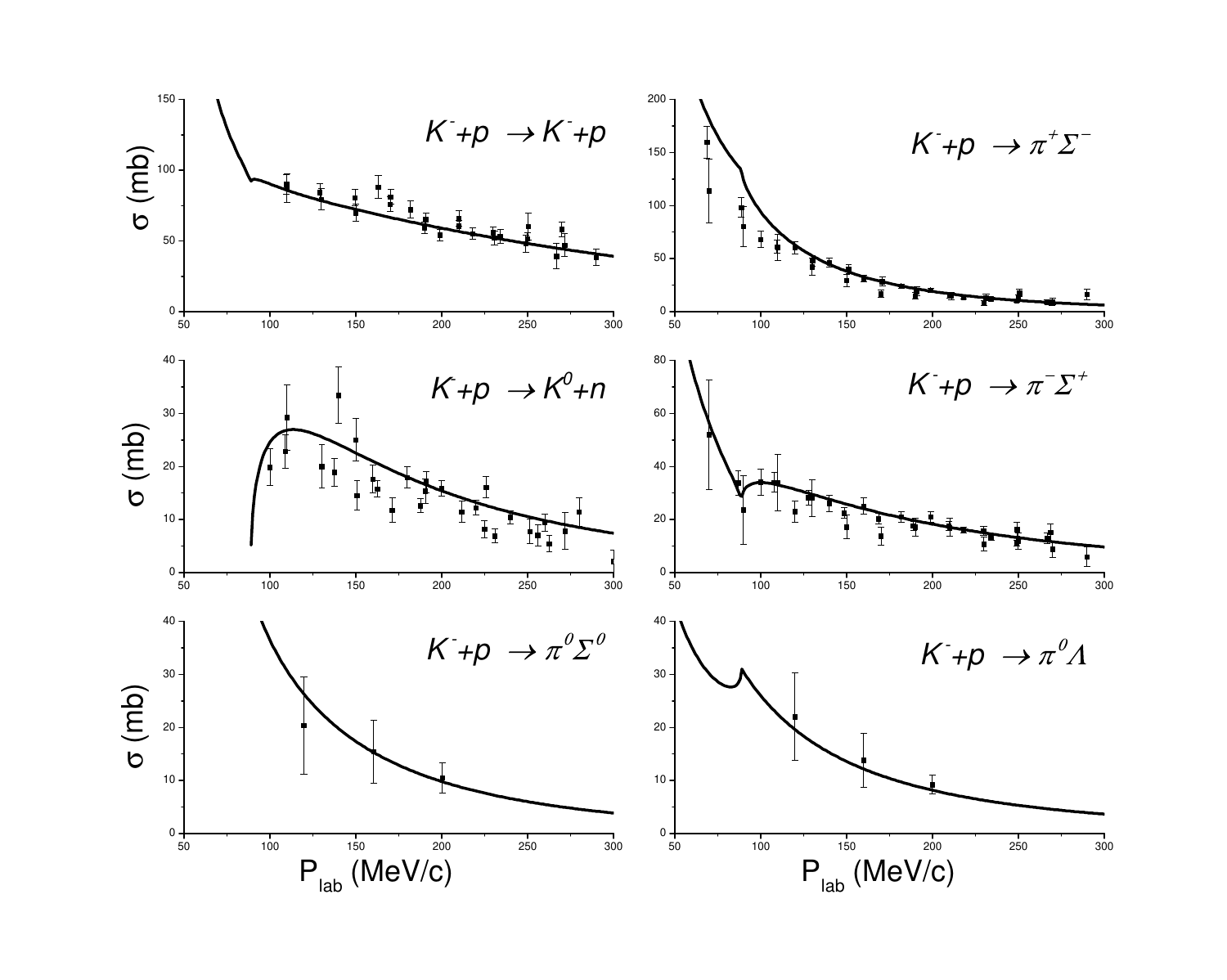}}
  \subfloat[Fit to CLAS data]{
    \includegraphics[width=0.485\linewidth]{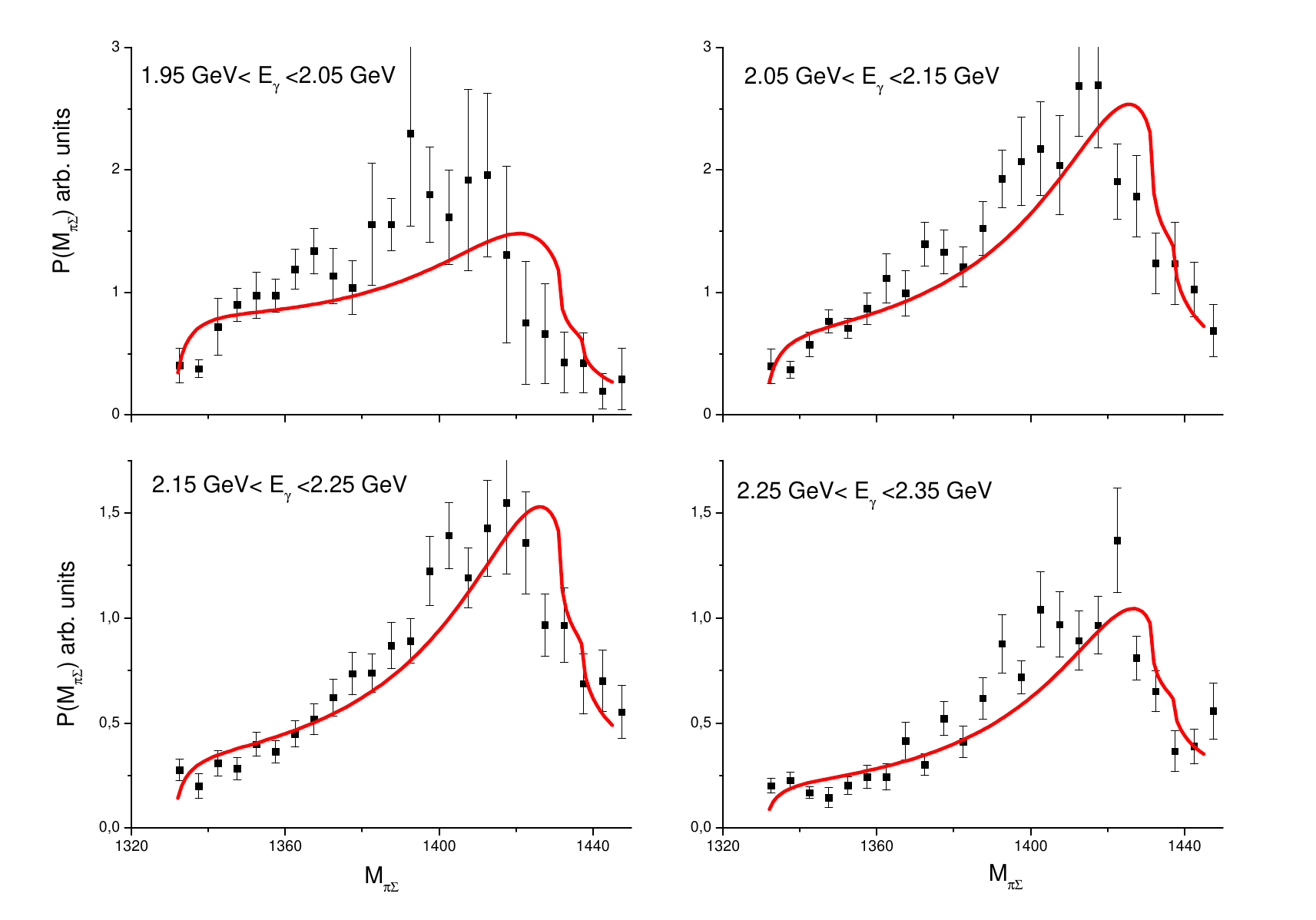}}
  \caption{Results of fit \bf{B}}
  \label{B}
\end{figure}

In view of this inability of the simple WT-like potential to produce simultaneously reasonable values both for $E(\Lambda^*)$ and $\Delta E$, we tried to reach a compromise by adding to the experimental data an artificial $E(\Lambda^*_c)$ somewhere below the $\bar{K}N$ threshold, say
\begin{equation}
E(\Lambda^*_c)=(1425\pm 5)-(25\pm 5)i\ MeV
\label{good}
\end{equation}
and performed the fitting including both $E(\Lambda^*_c)$ and $\Delta E$. This is our fit {\bf C}, the results of which are shown in column {\bf C} of tables I and II, and in part a) of Fig.3. The values of $E(\Lambda^*)_{\bf C}$ and $\Delta E_{\bf C}$ are acceptable as well as the quality of reproduction of "classical" data,  despite the fact, that  Re$(\Delta E_{\bf C})$
 is  somewhat above the experimentally allowed region.
We think, that for practical purposes, that is for calculation of few-body kaonic nuclear systems the choice {\bf C} is the most acceptable.

\begin{figure}
  \centering
  \subfloat[Fit to classical cross sections .]{
    \includegraphics[width=0.485\linewidth]{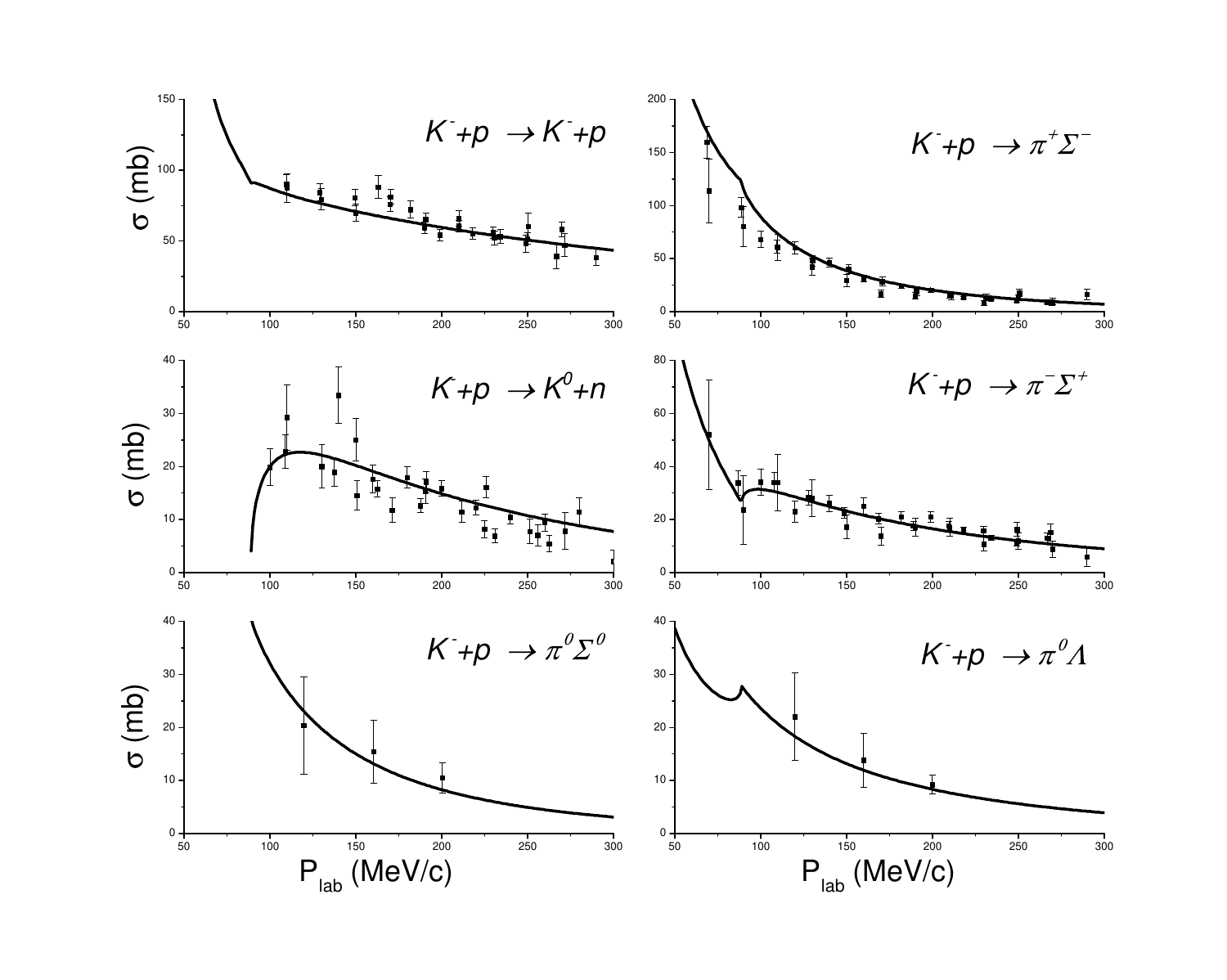}}
  \subfloat[Fit to CLAS data]{
    \includegraphics[width=0.485\linewidth]{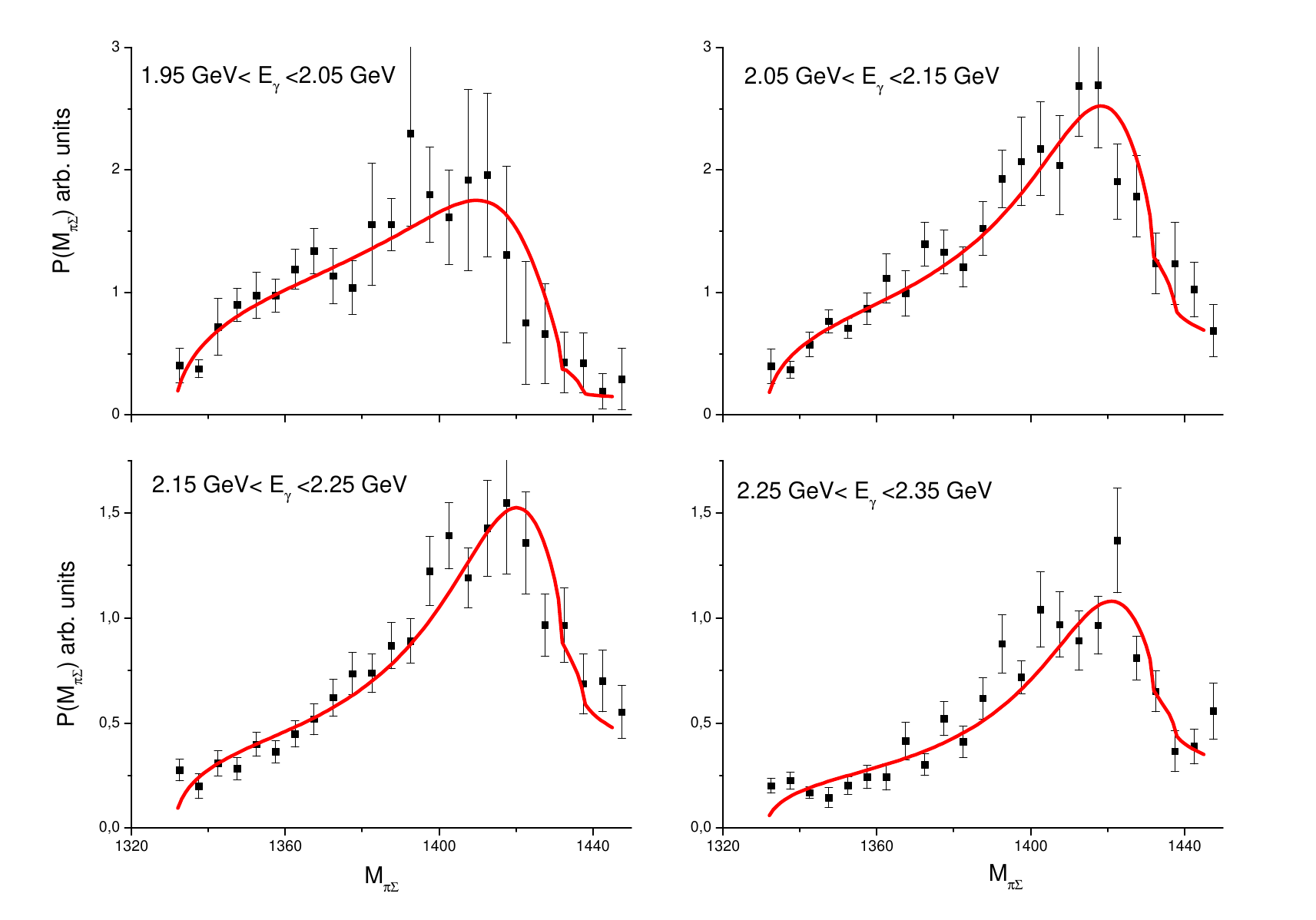}}
  \caption{Results of fit \bf{C}}
  \label{C}
\end{figure}

We have made two more fits to demonstrate the effect of strong $\bar{K}N$ binding on data reproduction. For this purpose we have added to the experimental data an $E(\Lambda^*)$ practically coinciding with the PDG value with somewhat increased uncertainties
to leave room for the fit:
\begin{equation}
E(\Lambda^*)=(1405\pm 2)-(25\pm 2)i\ MeV.
\label{PDG}
\end{equation}

In fit {\bf D} the $E(\Lambda^*)$ of (\ref{PDG}) was combined with the classical data, while in fit {\bf E}
also the experimental $\Delta E$ was added. The results are displayed in columns {\bf D} and {\bf E} of tables I and II and in part a) of Figs. 4. and 5.  In these fits the very small errors in (\ref{PDG}) force the
pole of the resulting potential to be close to the PDG value (\ref{PDG}). The price is a drastic worsening of overall fit quality, expressed in more than 100-150 \% increase of $\chi^2$ as compared to earlier fits.
In view of the above said, we don't think, that the assumption of strong $\bar{K}N$ binding is compatible
with the existing experimental data.

\begin{figure}
  \centering
  \subfloat[Fit to classical cross sections .]{
    \includegraphics[width=0.485\linewidth]{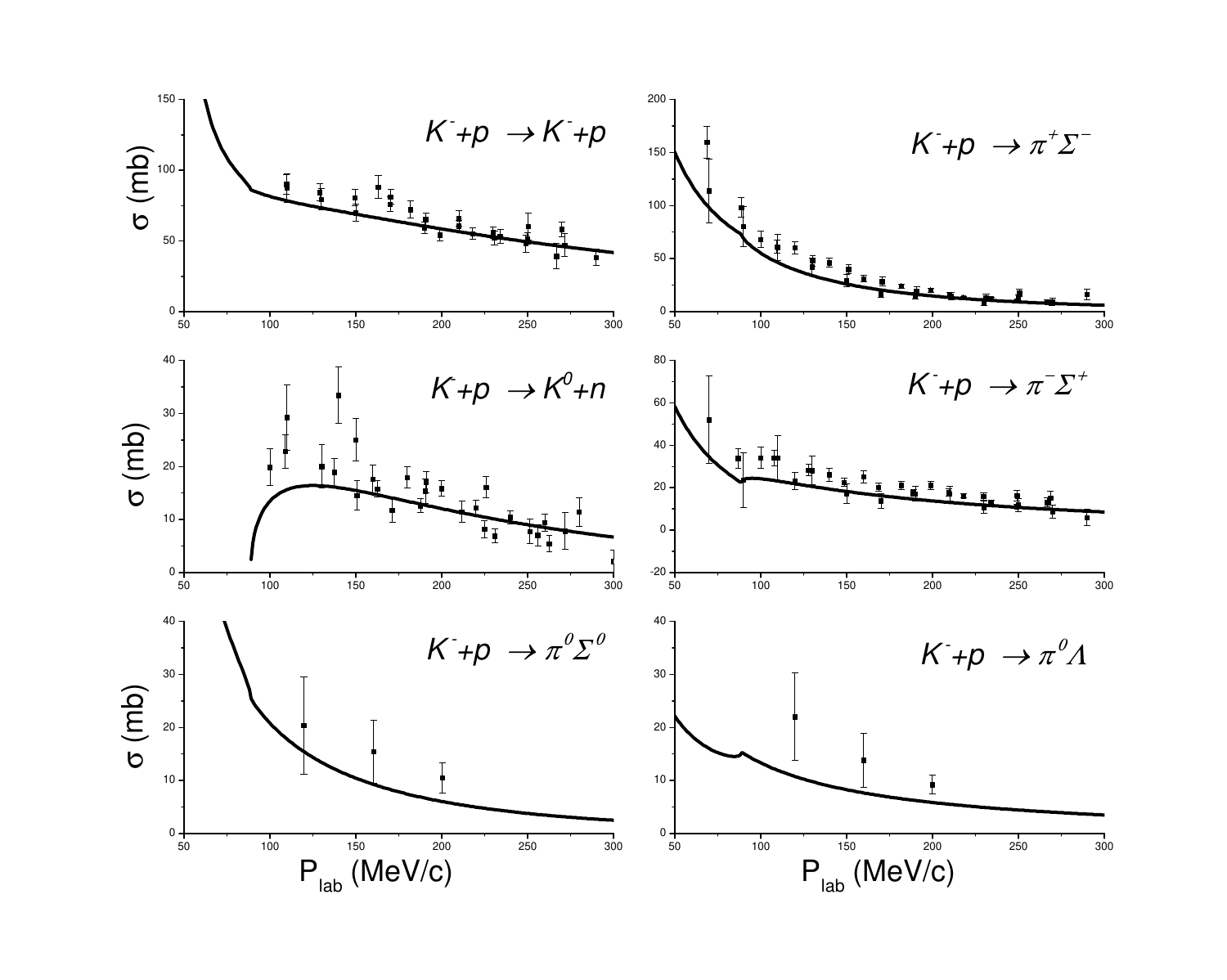}}
  \subfloat[Fit to CLAS data]{
    \includegraphics[width=0.485\linewidth]{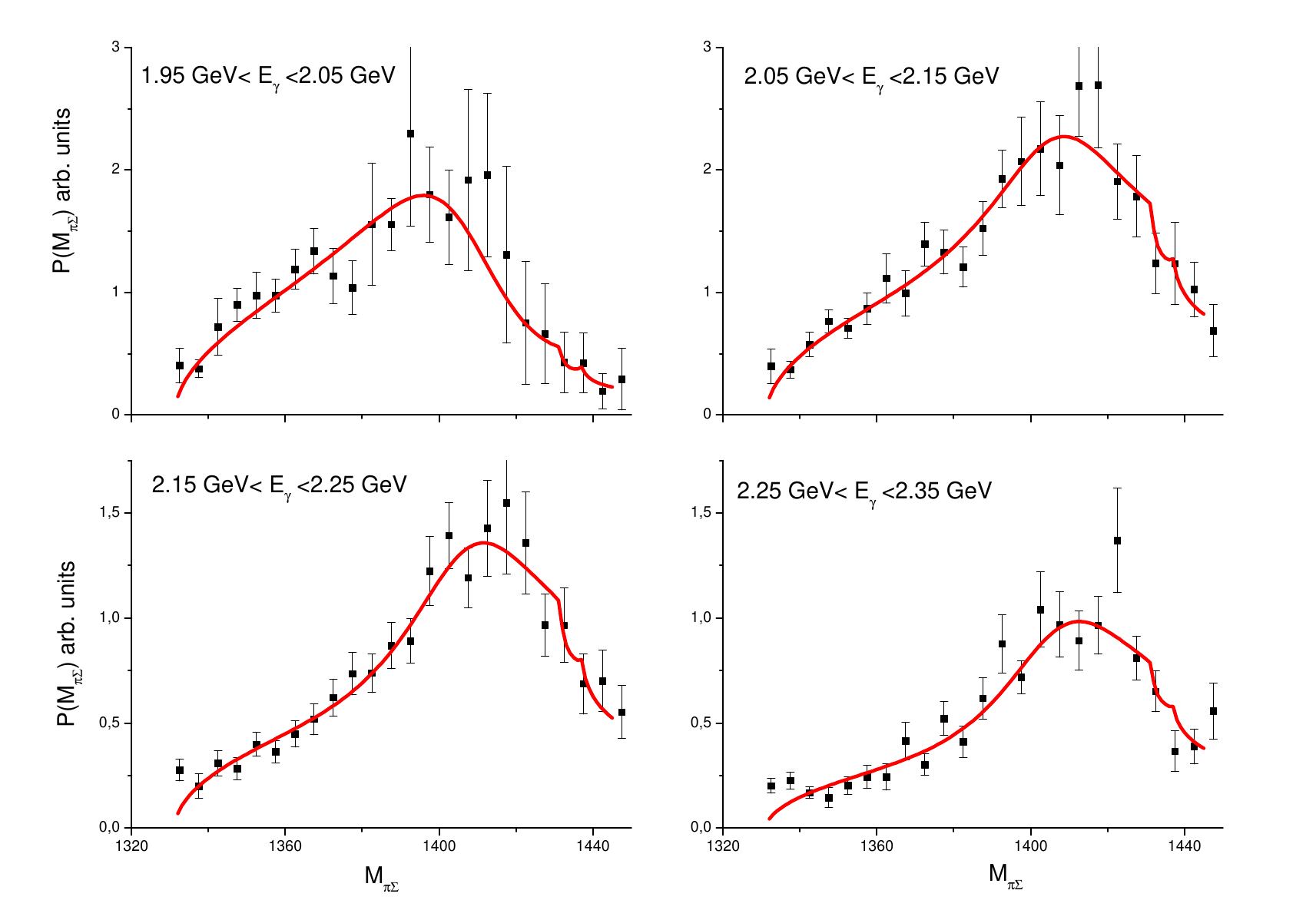}}
  \caption{Results of fit \bf{D}}
  \label{D}
\end{figure}

\begin{figure}
  \centering
  \subfloat[Fit to classical cross sections .]{
    \includegraphics[width=0.485\linewidth]{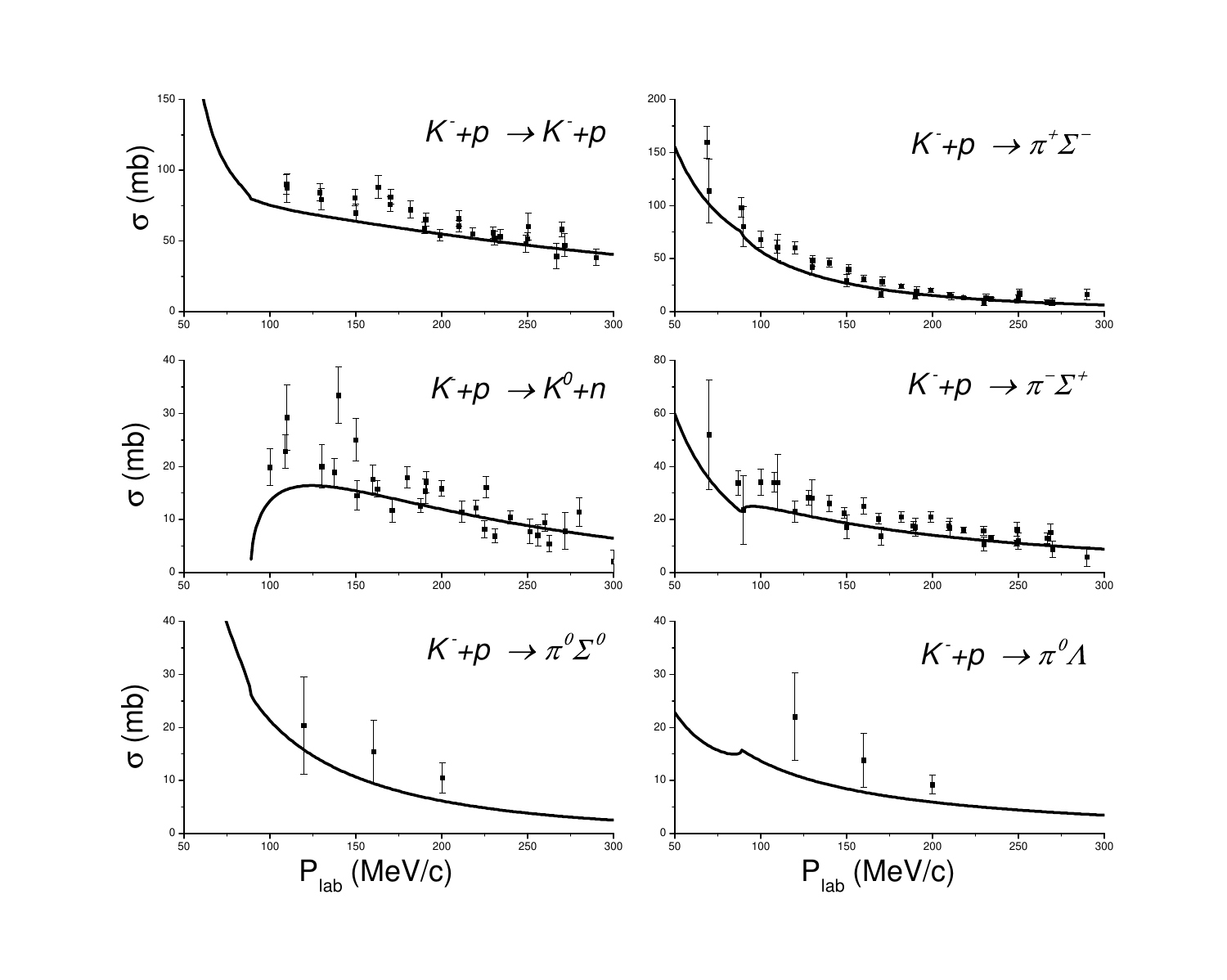}}
  \subfloat[Fit to CLAS data]{
    \includegraphics[width=0.485\linewidth]{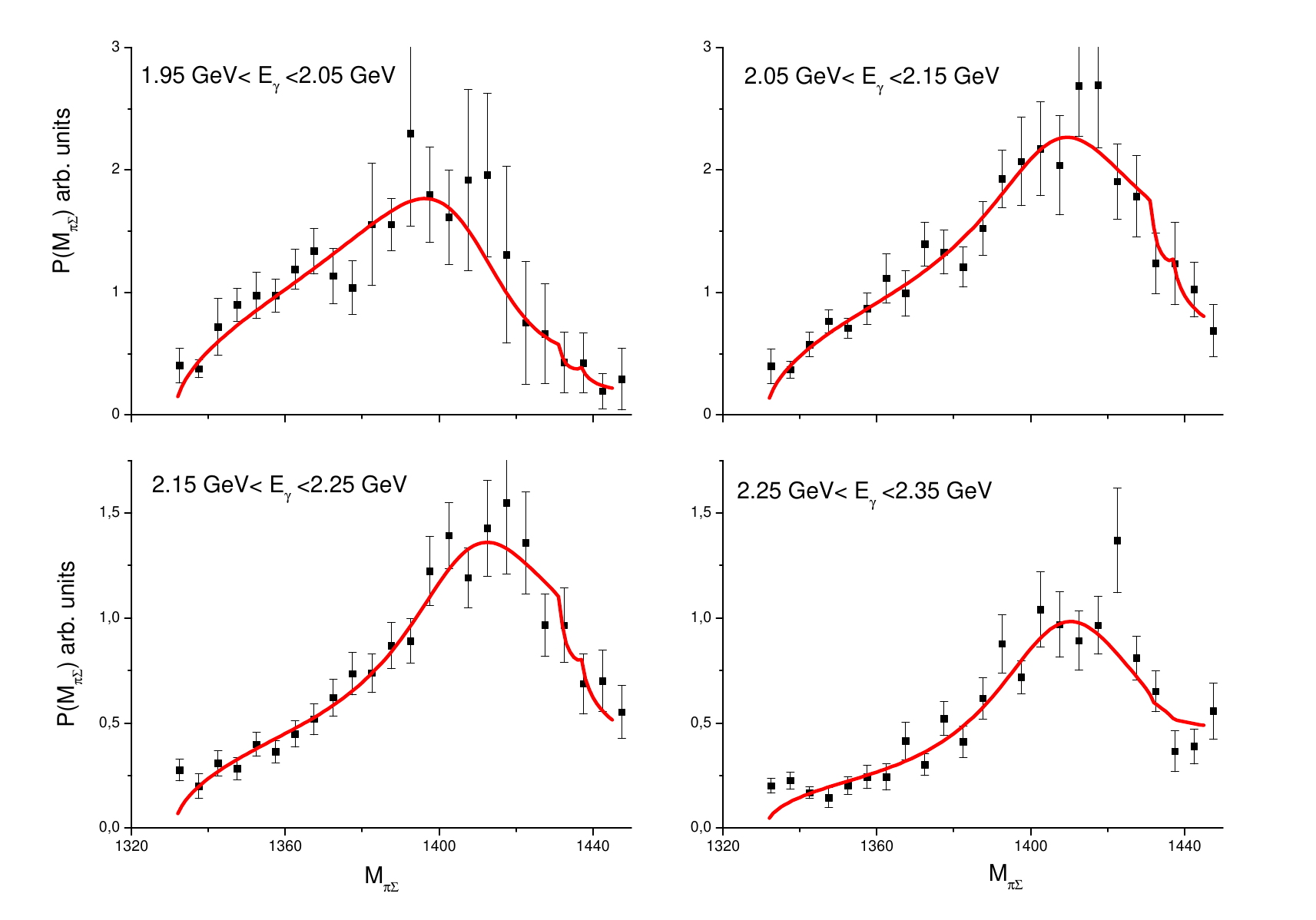}}
  \caption{Results of fit \bf{E}}
  \label{E}
\end{figure}

\section{Fitting to CLAS $\mathbf{\Lambda(1405)}$ data}

At this point it is reasonable to ask the question: how to incorporate the experimental information about the $\Lambda(1405)$ resonance into the fits, if not imposing an $E(\Lambda^*)$ pole position upon the $\bar{K}N$ interaction? We think, that the answer is: the fit should proceed through a theoretical description (model) - maybe approximate - of the reaction, where the resonance is produced. This description must contain in some way the $\bar{K}N$ interaction, about which one can make then conclusions by comparing the numbers produced by the model with the experimental data.

The most recent and accurate experimental information about the $\Lambda(1405)$ resonance comes from the recent CLAS photoproduction experiment \cite{CLAS}. In our opinion, from the abundance of CLAS data the best candidate for tracing the $\Lambda(1405)$ is the $\pi^0\Sigma^0$ missing mass spectrum from the reaction
\begin{equation}
\gamma+p\to K^++\pi^0+\Sigma^0
\label{photo}
\end{equation}
In this case the neutral $\pi$ and $\Sigma$ are in pure $I=0$ final state corresponding to $\Lambda(1405)$, while the charged final states contain not easily removable $I=1$ contributions, including $p$-waves from the $\Sigma(1385)$.

\begin{figure*}
\centering
\includegraphics[width=.6\linewidth]{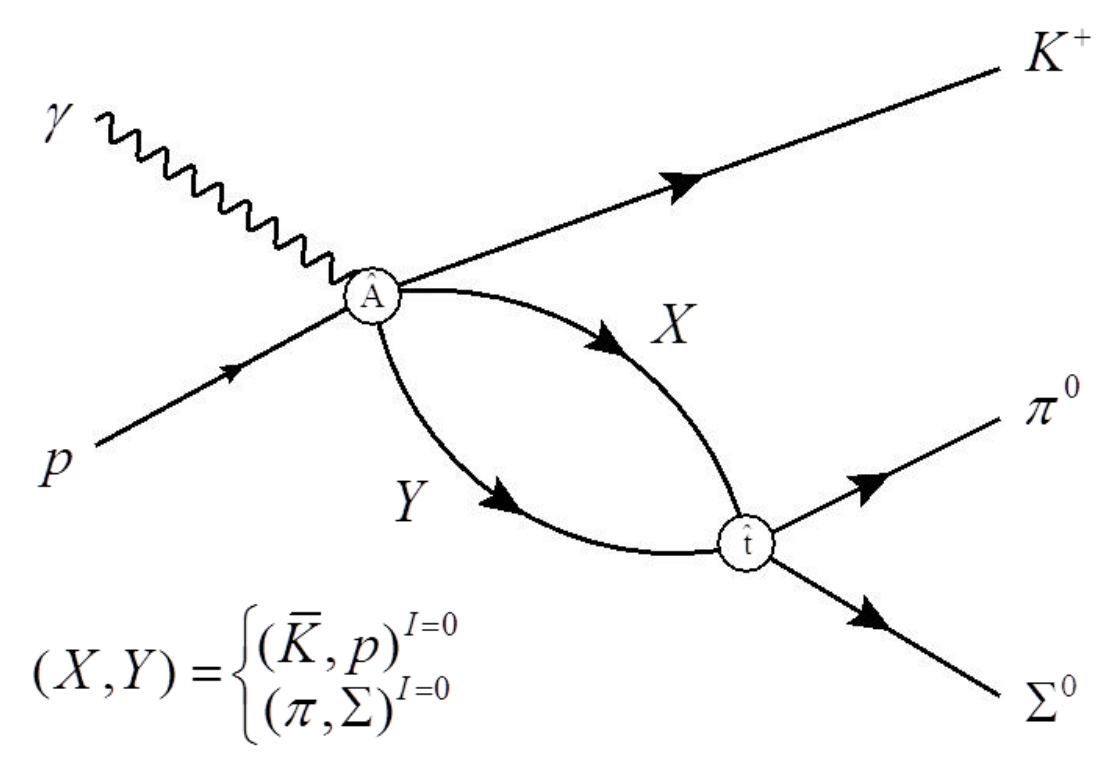}
\caption{Two-step model of \cite{Roca_Oset} for the reaction (\ref{photo}).}
\label{feyn}
\end{figure*}

For the reaction (\ref{photo}) Roca and Oset  proposed \cite{Roca_Oset} a two-step model corresponding to the diagram on Fig.\ref{feyn}, which was also applied in \cite{Mai_Meiss}. The corresponding amplitude can be written (somewhat schematically) as:
\begin{equation}
\begin{gathered}
\langle\mathbf{p}_{\gamma}|\hat{A}\hat{G}_0(E)\hat{t}|\mathbf{p}_{K^+},\mathbf{p}_{\pi^0},\mathbf{p}_{\Sigma^0}\rangle=\\
\sum_{(X,Y)}\int \langle\mathbf{p}_{\gamma}|\hat{A}|\mathbf{p}_{K^+},\mathbf{p}_X,\mathbf{p}_Y\rangle G_0(p_{K^+},p_X,p_Y;E)
\langle\mathbf{p}_X,\mathbf{p}_Y|\hat{t}|\mathbf{p}_{\pi^0},\mathbf{p}_{\Sigma^0}\rangle d\mathbf{p}_Xd\mathbf{p}_Y
\end{gathered}
\end{equation}

For fixed $\mathbf{p}_{\gamma}$ and $\mathbf{p}_{K^+}$ momentum conservation reduces the above integral to a single one over the relative momentum of particles $X,Y$:
\begin{equation}
%\begin{gathered}
f(k_{\pi^0\Sigma^0})=%\\
\sum_{(X,Y)}\int\langle\mathbf{p}_{\gamma}|\hat{A}|\mathbf{p}_{K^+},\mathbf{P}_{XY},\mathbf{k}_{XY}\rangle
G_0(p_{K^+},P_{XY},k_{XY};E)\langle k_{XY}|\hat{t}|k_{\pi^0\Sigma^0}\rangle d\mathbf{k}_{XY},
%\end{gathered}
\label{ampl}
\end{equation}
where the non-relativistic propagator $G_0(p_{K^+},P_{XY},k_{XY};E)$ is
$$
G_0(p_{K^+},P_{XY},k_{XY};E)=\left(E-m_{K^+}-m_{X}-m_{Y}-{p_{K^+}^2\over 2m_{K^+}}-{P_{XY}^2\over 2(m_X+m_Y)}
-{k_{XY}^2\over 2\mu_{XY}}+i\varepsilon\right)^{-1}
$$
Here we introduced the total and relative momenta $\mathbf{P}_{XY}$ and $\mathbf{k}_{XY}$ of particles $X$ and $Y$, and
$\mathbf{P}_{XY}$ is fixed by momentum conservation $\mathbf{P}_{XY}$=$\mathbf{p}_{\gamma}-\mathbf{p}_{K^+}$. The last factor in (\ref{ampl}) is the half off-shell $t$-matrix element of our $\bar{K}N$ potential.

These unsophisticated formulae were presented to be able to point out an important difference of the present calculation and the original method used in \cite{Roca_Oset}. The basic assumption, which makes the model calculable, is that we can neglect the $\mathbf{k}_{XY}$ dependence of the unknown amplitude $\langle\mathbf{p}_{\gamma}|\hat{A}|\mathbf{p}_{K^+},\mathbf{P}_{XY},\mathbf{k}_{XY}\rangle$ and  treat it as a complex energy- and (X,Y)-dependent parameter to be fitted to experimental data. Apart from this assumption in \cite{Roca_Oset} the remaining integral of the propagator and the $t$-matrix element was evaluated using the so-called "on-shell factorization", which reduces the effect of propagation between the two steps of the process to a multiplication by an energy-dependent constant instead of integration. While accepting the basic assumption, in the present calculation the remaining integration was performed exactly. In \cite{Nakamura} an attempt was made to propose a calculable model for the amplitude
$\langle\mathbf{p}_{\gamma}|\hat{A}|\mathbf{p}_{K^+},\mathbf{P}_{XY},\mathbf{k}_{XY}\rangle$
of Eq.(\ref{ampl}), however reasonable fits to the CLAS data could be achieved only at the price of introducing
a number of new adjustable parameters.

Thus finally our fitting procedure to the CLAS $\pi^0\Sigma^0$ missing mass spectrum for a given $\gamma$ energy $E_{\gamma}$ involves two complex constants corresponding to the two possible intermediate states $(X,Y)$, while for the $t$-matrix elements we used our previous fits {\bf A}-{\bf E} without further adjustment. The actual number of fitted  parameters is three, since an overall phase factor of the amplitude cancels out from the cross section. In part b) of the Figs.[\ref{A}]-[\ref{E}] we have shown the fits to the four lowest $\gamma$ energy bins of the CLAS photoproduction data.
The lowest $\gamma$-energies and the low-energy part of the missing mass spectrum were chosen due to the non-relativistic nature of our formulation.

Having in mind the approximate nature of the applied reaction model, the quality of the fits seem to be acceptable, confirming  the final state interaction nature of the $\Lambda(1405)$ peak. The only exception is fit {\bf B}, where the
$E(\Lambda^*)$ pole position moved above the $\bar{K}N$ threshold. This case is unambiguously ruled out by the CLAS data. It is interesting to note, that fits {\bf D} and {\bf E} with PDG values for $E(\Lambda^*)$, which yield poor results for the classical two-body data, still can be adjusted to reproduce the observed peaks. This fact can be misleading if a potential is constructed mainly to fit the observed $\Lambda(1405)$ line shapes.

\section{About the acceptance of the potential}
Before drawing our conclusions concerning the fits with this potential, it could be appropriate to comment on two papers  \cite{MY,BC}, criticizing it.
In both papers the starting point is the same as in \cite{revai1}: the lowest order Weinberg-Tomozawa (WT) term of the chiral expansion of the SU(3) meson-baryon interaction Lagrangian.

In the first one (\cite{MY}) mainly the statement "the two-pole structure of the $\Lam$ is due to the use of the on-shell approximation" is challenged, while it did not question the applicability or usefulness of the potential in non-relativistic calculations. The authors presented a two-channel $\bar{K}N$ calculation  using relativistic kinematics and dimensional regularization of the divergent integrals with and without the on-shell approximation and ,ideed, in both cases they found two poles in the $I=0$ sector. No comparison with experimenatal data was presented. It is interesting to note, that the poles calculated without the on-shell approximation lie much further away from the physical region, than those obtained by using it.

The second paper (\cite{BC}) consists of two parts. In the first one the potential \cite{revai1} is scrutinized remaining in the non-relativistic framework. While confirming the obtained results, it is found, that the chiral limit of the meson-baryon scattering length, derived from the SU(3) effective field theory, is not reproduced by this potential. It is discussed in some detail, why this feature is a serious drawback of the potential, without pointing out, though, in wich respect this fact hinders the use of the potential in the physical, non-zero mass case for the purpose it was designed.
It has to be noted, that an {\it ab ovo} non-relativistic approach, which is based on the assumption of smallness of the relevant momenta in the process compared to the particle masses is not really compatible with the chiral limit, in which the zero mass mesons certainly do not satisfy this condition. The simultaneous application of both limits to a system or process does not seem to be reasonable. One can, of course, blame the non-relativistic approach, however, if the potential is basically designed for calculations in $n>2$ systems, where relativistic treatment is not really available, then the fitting of potential parameters to the two-body data has to be performed under the same conditions
In the second part of \cite{BC} the authors propose an extension of the disputed potential using relativistic kinematics without the on-shell approximation combined with a separable potential type regularization based on cut-off (form factor) functions. Keeping the number of adjustable parameters as low as possible - and consequently a not too convincing - fit to experimantal data was performed and, indeed, two poles on the unphysical $\pi\Sigma$ sheet were found. 

In both relativistic realizations of the off-shell WT term one pole was found more or less close to the $\Lam$ position, while the second ones are very far both from each other  and from the physical $\Lam$. 

In view of the above said, the statement of \cite{revai1} about the origin of two pole structure of $\Lam$ has to be refined. Its validity is most likely restricted to the non-relativistic case, while calculations with relativistic kinematics can produce two poles even with off-shell WT interaction. 

Finally, let us formulate our opinion in the long standing debate on the one- or two-pole nature of $\Lam$. 

In this context the appearance of a very detailed analysis of all available experimental data in a somewhat extended energy range together with a comprehensive 
compilation of the $\Lam$ related literature has to be mentioned \cite{hyperonIII}. From our point of view this analysis is a polite rejection of any relevance of the second pole in the reproduction of data. Perfect fits for all kind of data are achieved assuming only one pole in the $\Lam$ region 
(it turns out to be very close to the one predicted by \cite{revai1,revai2}), while imposing a two-pole scenario does not improve the fits and yields a second pole far in the non-physical region.  

Therefore, as we see it now, the physically observed $\Lam$ state can be associated with a single pole in the complex energy plane. The two-pole structure is a feature not of this physical state, but rather of certain realizations of the chiral SU(3) model of meson-baryon interaction. Confrontation of these model calculations with experimental data yield one pole reasonably close to the $\Lam$ position, while the position of the second ones is strongly model dependent and usually far from the $\Lam$ region. It is hard to define to what extent and in which sense these second poles are part of the physical state represented by the experimental resonance. In view of the above said we find it somewhat regrettable, that the categorical opinion about the two-pole structure of $\Lam$ penetrated also the PDG Review of Particle Physics \cite{PDG}. 

\section{Conclusions}

We have confronted a certain potential from class b) (see Introduction) with different pieces of experimental data.

Comparison with classical data set yields a good agreement with minimal $\chi^2$. From the unfitted quantities $E(\Lambda^*)$ shows a modest binding ($\approx 10\ MeV$) and a reasonable width ($\approx 40\ MeV$). On the other hand, the $1s$ level shift $\Delta E$ is far from its experimental value, its real part is strongly overestimated, while the imaginary part is just inside the experimentally allowed region.

When the experimental $\Delta E$ is added to the classical data, the resulting potential produces a $\Delta E$ with a real part close to the experimental value, while the imaginary part is slightly above the allowed region. The price of bringing the real part to the right place is, however, that $E(\Lambda^*)$ moved above the $\bar{K}N$ threshold, what contradicts to the $\Lambda(1405)$ experiments.

This sharp contradiction between the $E(\Lambda^*)$ position and experimental $\Delta E$, probably characteristic for most of the class b) potentials, can be smoothed by adding to the fit a requirement of having an  $E(\Lambda^*)$ "somewhere below the
$\bar{K}N$ threshold", say at $E(\Lambda^*_c)$ of (\ref{good}). The potential from the resulting fit {\bf C} can be considered as an acceptable compromise between the two extremes.

Finally, fits with $E(\Lambda^*)$ pinned down close to the PDG value produce unacceptable results both for the classical data set and $\Delta E$, while they are able to reproduce the peaks in the photoproduction experiment.


\begin{thebibliography}{1}

\bibitem{Mai}
M. Mai, Status of the $\Lam$,  Few-Body Syst. {\bf 59},61(2018)

\bibitem{Kamiya}
Y.Kamiya et al., Antikaon–nucleon interaction and $\Lam$ in chiral SU(3) dynamics, Nucl. Phys. A {\bf 954},41(2016)

\bibitem{Cieply}
A. Cieply et al., On the pole content of coupled channels chiral approaches used for the $\bar{K}N$ system, Nucl. Phys. A {\bf 954},17(2016)

\bibitem{revai1}
J. R\'{e}vai, Are the chiral based $\bar{K}N$ potentials really energy-dependent?, Few-Body Syst. {\bf 59},49(2018)

\bibitem{revai2}
J. R\'{e}vai, Energy Dependence of the $\bar{ K}N$ interaction and the two-pole structure of the $\Lam$ -- are they real?,
N. A. Orr et al. (eds.), Recent Progress in Few-Body Physics,Springer Proceedings in Physics 238,p. 947, Springer Nature Swityerland AG,2020

\bibitem{Guo}
Z.-H. Guo, J. A. Oller, Meson-baryon reactions with with strangeness -1 within a chiral framework,  Phys. Rev. C {\bf 87}, 035202(2013)

\bibitem{expref}
D.N. Tovee et al., Some properties of the charged $\Sigma$ hyperons, Nucl. Phys. B {\bf 33}, 493 (1971)
 
 R.J. Nowak et al., Charged $\Sigma$  hyperon production by $K^-$ meson interactions at rest, Nucl. Phys. B {\bf 139}, 61 (1978)
 
 M. Sakitt et al., Low-energy $K^-$-meson interactions in hydrogen, Phys. Rev. B {\bf 139}, 719 (1965)
 
 J.K. Kim, Low-energy $K^-p$ interaction and interpretation of the 1405-MeV $Y_0^*$ resonance as a $\bar{K}N$ bound state, Phys. Rev. Lett. {\bf 14}, 29 (1965)

J.K. Kim, Multichannel phase-shift analysis of $\bar{K}N$ interaction in the region 0 to 550 MeV/c, Phys. Rev. Lett. {\bf 19}, 1074(1967)

W. Kittel, G. Otter, I. Wacek, The $K^-$  proton charge exchange interactions at low energies and scattering lengths determination, Phys. Lett. {\bf 21}, 349 (1966)

J. Ciborowski et al., Kaon scattering and charged Sigma hyperon production in $K^-p$ interactions below 300 MeV/c, J. Phys. G {\bf 8}, 13 (1982)

D. Evans et al., Charge-exchange scattering in $K^-p$ interactions below 300 MeV/c, J. Phys. G {\bf 9}, 885 (1983)

\bibitem{SIDD}
M. Bazzi et al. (SIDDHARTA Collaboration), A new measurement of  kaonic hydrogen X-rays, Phys. Lett. B {\bf 704},113(2011)

\bibitem{Bora}
B. Borasoy, U.-G. Meissner, R. Nissler, $K^-p$ scattaring length from scattering experiments, Phys. Rev. C {\bf 74},055201(2006)

\bibitem{Ikeda}
Y. Ikeda et al. , Chiral SU(3) theory of antikaon–nucleon interactions with improved threshold constraints, Nucl. Phys. A {\bf 881},98(2012)

\bibitem{CLAS}
K. Moriya et al.(CLAS Collaboration), Measurement of the $\pi\Sigma$ photoproduction line shapes near the $\Lam$, Phys. Rev. C {\bf 87},035206(2013)

\bibitem{Roca_Oset}
L. Roca and E. Oset, $\Lam$ poles obtained form $\pi^0\Sigma^0$ photoproduction data, Phys. Rev. C {\bf  87},055201(2013)

\bibitem{Mai_Meiss}
M. Mai, U.-G. Meissner, Constrains on the chiral unitary$\bar{K}N$ amplitude from $\pi\Sigma K^+$ photoproduction data, Eur.  Phys.J. A {\bf 51},30(2015)

\bibitem{Nakamura}
S. X. Nakamura and  D. Jido, Lambda (1405) photoproduction based on the chiral unitary model,  Progr. Theor. Exp. Phys. {\bf 2014},023D01

\bibitem{MY}
O. Morimatsu and K. Yamada, Renormalization of the unitarized Weinberg-Tomozawa interaction without on-shell factorization and $I=0$ $\bar{K}N-\pi\Sigma$ coupled channels, Phys.Rev. C {\bf 100},023201(2019)

\bibitem{BC}
P.C. Bruns and A. Cieply, Importance of chiral constraints for the pole content of the $\bar{K}N$ scattering amplitude, Nucl. Phys. A {\bf 996},121702(2020) 
 
\bibitem{hyperonIII}
A. V. Anisovich  et al., Hyperon III:$K^-p\mbox{ -- }\pi\Sigma$ coupled-channel dynamics in the $\Lam$ mass region, Eur. Phys. J. A {\bf 56},139(2020)

\bibitem{PDG}
U-G. Meissner and T. Hyodo, 101.Pole structure in the $\Lam$ region,  M. Tanabashi et al. (Particle Data Group), Phys. Rev. D {\bf 98}, 030001 (2018) and 2019 update













\end{thebibliography}
\end{document}